\documentclass[conference]{IEEEtran}
\pdfoutput=1
\usepackage[english]{babel}
\usepackage[utf8]{inputenc}
\usepackage{amsmath}
\usepackage{graphicx}
\usepackage{float}
\usepackage{hyperref}
\usepackage{cite}
\usepackage{color}

\definecolor{green}{RGB}{0,190,0}
\definecolor{grey}{RGB}{128,128,128}
\newcommand\numberthis{\addtocounter{equation}{1}\tag{\theequation}}

\begin{document}
\title{PageRank Approach to Ranking National Football Teams}

\author{\IEEEauthorblockN{Verica Lazova}
\IEEEauthorblockA{Faculty of Computer Science and Engineering\\
Cyril and Methodius University\\
Skopje, R.Macedonia\\
lazova992 at gmail.com}
\and
\IEEEauthorblockN{Lasko Basnarkov}
\IEEEauthorblockA{Faculty of Computer Science and Engineering\\
Cyril and Methodius University\\
Skopje, R.Macedonia\\}}


\maketitle

\begin{abstract}
The Football World Cup as world's favorite sporting event is a source of both entertainment and overwhelming amount of data about the games played. In this paper we analyse the available data on football world championships since 1930 until today. Our goal is to rank the national teams based on all matches during the championships. For this purpose, we apply the PageRank  with restarts algorithm to a graph built from the games played during the tournaments. Several statistics such as matches won and goals scored are combined in different metrics that assign weights to the links in the graph. Finally, our results indicate that the Random walk approach with the use of right metrics can indeed produce relevant rankings comparable to the FIFA official all-time ranking board.

\end{abstract}

\IEEEpeerreviewmaketitle

\section{Introduction}
Football, being the world's most favored sport, draws people's attention in every field, from the simple means of entertainment to more complex objectives of statistics, research and data analysis. Since the FIFA world cup first took place in 1930 until this day, there have been around 20 tournaments held, each comprising of about 64 matches, not counting the qualification rounds \cite{fifa-competitions, fifa-comparative}. Therefore, there is significant amount of data that one could inspect, analyse and draw conclusions from. 

Having that in mind researchers are tackling problems regarding playing strategy, ranking of teams or performance analysis from different aspects including economic, demographic, cultural and climatic factors \cite{hoffmann2002socio}. A team's game strategy for example can be observed from graph theory perspective by constructing a network of passes between players.
In this context different centrality measures can be used to determine the importance of particular players \cite{pena2012network, duch2010quantifying, hughes2005analysis}. Other subject of interest might be modelling football matches in terms of scores during the game. For example, in \cite{dixon1998birth} the authors discuss a statistical model for scoring times in a match.

Here we address the problem of ranking national football teams. Our main task is to use the available statistics, in order to come up with an alternative ranking method for the football teams based on their achievements at the world cups. There are different rating methods currently in use and they produce relevant results. FIFA have their own 4-year points based FIFA/Coca-Cola rating system \cite{fifa-coca-cola} and world cup all-time ratings \cite{fifa-all-time-rankings} that includes all championships since their origin. There are also the World Football Elo Ratings based on the rating system FIDE uses to rate chess players~\cite{elo}.  

A good ranking method should not only take into account how many times a team has won, but also consider how strong an opponent they have defeated. Victory against stronger opponent is preferable and thus more significant than victory against weaker opponent. One method that incorporates such logic is the PageRank (Random walk) method, which is applicable to vast varieties of network based problems that require ranking in some way. Other than the well known problem of rating web-pages  \cite{page1999pagerank} it is also utilized in social network analysis, in tasks such as link prediction, information diffusion and communities detection \cite{backstrom2011supervised, kimura2006tractable, stanoev2011identifying}. Also it is used in  NLP for the purpose of text summarization and word sense disambiguation \cite{erkan2004lexrank, agirre2009personalizing}. For previous attempts of employing PageRank mechanism in sporting events we refer the reader to \cite{keener1993perron, mukherjee2012identifying, radicchi2011best}. 

The rest of the paper is organized as follows. In Section~\ref{s2} we present the ranking problem and the PageRank based method for solving it. We also give description and statistics of the data that was available to us. The obtained results are presented in Section~\ref{s3} including a discussion and comparison to the official rankings and then we conclude the paper in Section~\ref{s4}.

\section{Materials and Methods}
\label{s2}

\subsection{Data}

The data we used was obtained from 11v11, web-site for football statistics that contains all time figures about the matches of the world cup, qualification games inclusive~\cite{11v11}. For each national team there is information on which country they have played against, the number of matches won, drawn and lost, as well as the number of scored and conceded goals during all match-ups. Throughout this paper we use the term \textit{match-up} in context of \textit{a single game played between two teams}. And a \textit{match-up pair} are \textit{every two teams that have played against each other}. 
	The dataset, contains 210 countries and statistics on 2335 match-up pairs that have played against one another, or 7141 games in total, during which 20298 goals were scored. The average number of games per match-up pair is  3.0582, and the average number of goals scored per match-up pair is 4.3465. Mexico versus USA is the pair with the largest number of games played against one another. About 28 games were played during which around 100 goals were scored, 15 of which were won by the US, 6 were drawn and the other 7 resulted in a victory for Mexico. The country with the most games played is Brazil with about 200 matches and also is the country with most games won and most goals scored as expected. 

\subsection{Method}
The ranking method explored throughout this paper is the PageRank with restarts algorithm applied to a graph build around the supplied data~\cite{page1999pagerank}. Each national team is a single node in the graph and two nodes are linked if the two teams (the match-up pair) have ever competed against each other in a world cup tournament. The weight of the link is determined by a weighting function that involves one or more metrics such as number of games played between a match-up pair, the number of won, lost and drawn games, or the number of scored and conceded goals. The various weighting functions we have tested are given in Table~\ref{tab1}. 

\renewcommand{\arraystretch}{1.5}
\begin{table}
\scriptsize
\centering
\caption{\label{tab1} Set of tested weighting function and their score in normalized number of inversions as similarity metrics to the official rankings. Less is better.}
\begin{tabular}{| c | l | c |}
\hline
\textbf{\#} & \textbf{WEIGHTING FUNCTION} & \textbf{INVERSIONS} \\ \hline
1  & \(f_{i,j} = \frac{l_{i, j}}{g_{i,j}} \cdot \frac{1}{G-g_{i,j}+1}\) & 0.032\\ \hline
2  & \(f_{i,j} = \frac{l_{i, j}}{g_{i,j}}\) & 0.038\\ \hline
3  & \(f_{i,j} = \frac{l_{i, j}}{g_{i,j}} + \frac{c_{i, j}}{c_{i, j} + s_{i, j}}\) & 0.040\\ \hline
4  & \(f_{i,j} = l_{i, j}\) & 0.040\\ \hline
5  & \(f_{i,j} = \frac{c_{i, j}}{s_{i, j}}\) & 0.041\\ \hline
6  & \(f_{i,j} = \frac{l_{i, j}}{w_{i, j}}\) & 0.043\\ \hline
7  & \(f_{i,j} = \frac{l_{i, j}}{g_{i,j}} + 0.5\cdot\frac{d_{i,j}}{g_{i,j}}\) & 0.044\\ \hline
8  & \(f_{i,j} = \frac{c_{i, j}}{c_{i, j} + s_{i, j}}\) & 0.044\\ \hline
9  & \(f_{i,j} = \frac{c_{i, j}}{g_{i,j}}\) & 0.046\\ \hline
10 & \(f_{i,j} = c_{i, j}\) & 0.050\\ \hline
\end{tabular}

\end{table}
Within the functions we use the following notation: 
\begin{description}
\item[\(f_{i, j}\)] weight of the link from node \(i\) to node \(j\);
\item[\(g_{i, j}\)] number of games played between the two teams;
\item[\(l_{i, j}\)] number of games lost by team \(i\) amongst all the games \(i\) and \(j\) played;
\item[\(w_{i, j}\)] number of games won by team \(i\) amongst all the games \(i\) and \(j\) played;
\item[\(c_{i, j}\)] number of goals conceded by team \(i\) during all the games \(i\) and \(j\) played;
\item[\(s_{i, j}\)] number of goals scored by team \(i\) during all the games \(i\) and \(j\) played;
\item[\(d_{i, j}\)] number of games drawn between the two teams;
\item[\(G\)] maximum number of games played between any match-up pair; 
\end{description}

Another factor that affects the PageRank is the damping factor. The damping factor corresponds to the probability that a random walker would discontinue the walk and jump to a random node~\cite{brin1998anatomy}. The damping factor other than being necessary as assurance that the random walk would converge to a stationary distribution, it is also intuitive. The intuition behind the use of damping factor within our match-ups network is the following: although the graph is dense not every team have played against every other. So when using weighting metrics such as the loss ratio (funcion 1 in Table~\ref{tab1}) the damping factor would mean adding some wining chances to all the teams that have never been played against. It also adds some wining chances to a team that has never won a game within a match-up.

The PageRank is calculated using the power method \cite{langville2004deeper}. 
This method is an iterative algorithm (eq.~\ref{eq2}) that finds the dominant eigenvector, which corresponds to the invariant distribution of the time a random walker spends at a certain node - the PageRank. 
By normalizing the adjacency matrix $A$ we get the transition probability matrix $Q$ with elements as given in eq.~\ref{eq1}.

\begin{align*}
Q_{i,j} &= (1-d) \cdot \frac{A_{i,j}}{\sum\limits_{k=1}^N A_{i,k}} + \frac{d}{N} \numberthis \label{eq1}\\ 
\pi^T &= \pi^TQ. \numberthis \label{eq2} 
\end{align*}

Note that $Q$ is guaranteed to be irreducible and aperiodic as a consequence of the nonzero damping factor $d$.


\begin{table}[h]
\centering
\scriptsize
\caption{\label{tab2} Number of games played and results for each match-up pair} 
\begin{tabular}{| l | c | c |} 
	\hline
	\textbf{PAIR} & \textbf{GAMES} & \textbf{RESULTS} \\ \hline
	A-B & 3 & A wins 2, B wins 1 \\ \hline
    A-C & 3 & A wins 2, C wins 1 \\ \hline
    A-D & 3 & A wins 3, D wins 0 \\ \hline
    B-C & 3 & C wins 3, B wins 0 \\ \hline
    B-D & 3 & D wins 3, B wins 0 \\ \hline
    C-D & 3 & C wins 1, D wins 2 \\ 
    \hline    
\end{tabular}

\caption{\label{tab3} The PageRank of each team in descending order} 
\begin{tabular}{| c | c | c | c | c |}
    \hline
    \textbf{TEAM} & \textbf{GAMES} & \textbf{WIN} & \textbf{PAGERANK} \\ 		\hline 
    A & 9 & 7 & 0.333 \\ \hline
    C & 9 & 5 & 0.281 \\ \hline
    D & 9 & 5 & 0.211 \\ \hline
    B & 9 & 1 & 0.175 \\ 
    \hline
\end{tabular} 
\end{table}

\begin{figure}[h]
\centering
\includegraphics[width=0.45\textwidth]{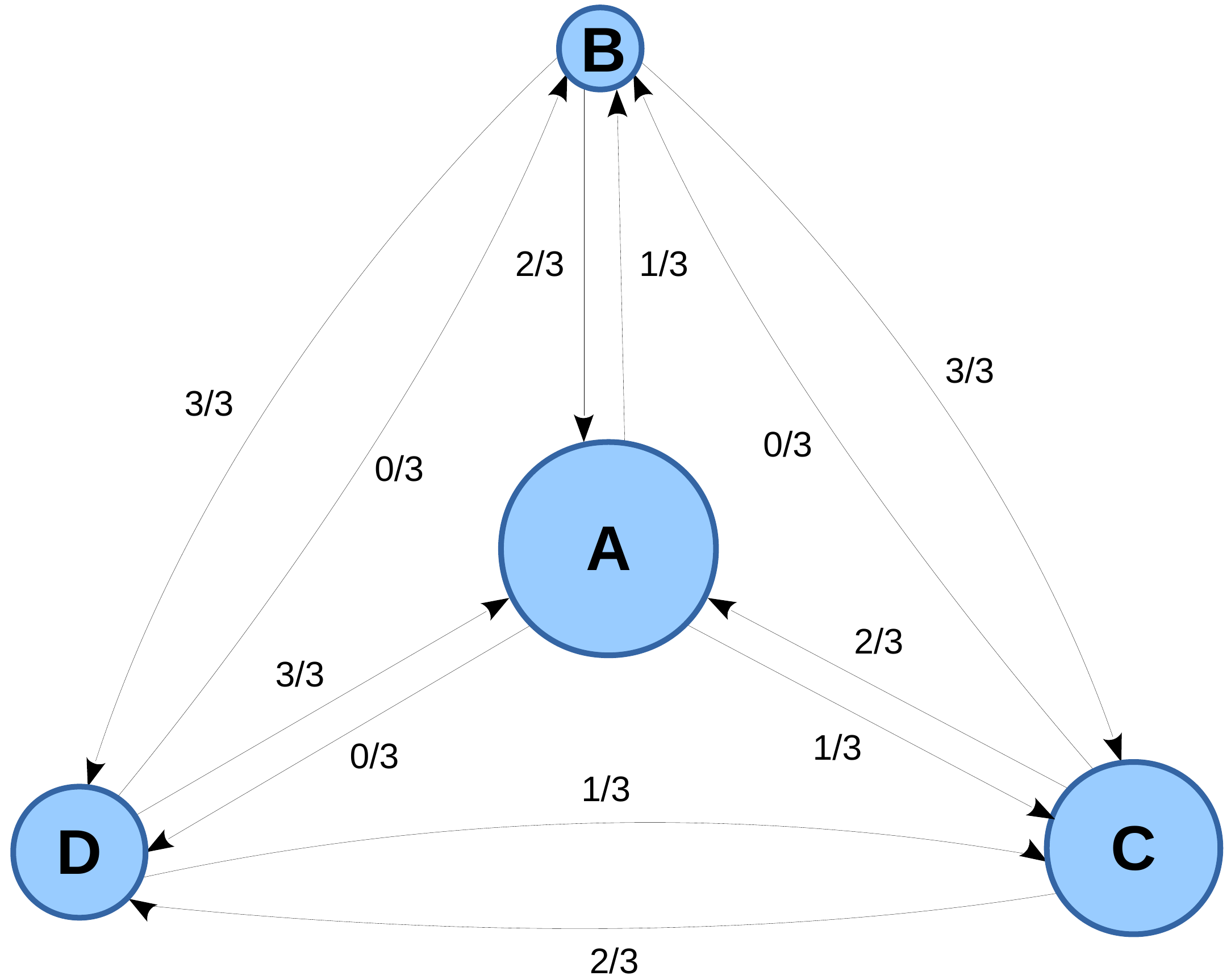}
\caption{\label{fig1} Graph representation og the games played, the size of each node is proportional to it's PageRank}
\end{figure}

\subsection{Example}

For the sake of demonstration, let's consider a toy example that illustrates our goal. Suppose there are 4 teams and the given statistics for each pair are shown in Table~\ref{tab2}. The graph (Fig.~\ref{fig1}) is built using loss ratio as metric (function 2 at Table~\ref{tab1}). Therefore the weight of a given link from $i$ to $j$ is the part of the games that $i$ has lost to $j$. For instance there is a link from A to C with weight of \(\frac{1}{3}\) and also a link from C to A with weight of \(\frac{2}{3}\). That means out of 3 matches A and C have played against each other A has won 2 matches, C has won 1 and no matches were drawn. The next step is calculation of the PageRank. Therefore we need transition probability matrix which is calculated according to eq.~\ref{eq1} with a common damping factor value of 0.15.

Finally the results are shown at Table~\ref{tab3}. A is pointed as highest ranked and B is lowest ranked team as expected. On the other hand, team C and team D both have won 5 games as shown in Table~\ref{tab3}. However, PageRank takes into account the strength of the defeated opponent not only the number of winnings. As a result, team C is ranked higher since they have won a game against A, considered as strong opponent, in contrast to team D who have winnings only against weaker opponents. 

\section{Results and Discussion}
\label{s3}

\begin{table}[h]
\centering
\scriptsize
\caption{\label{tab4} Top 20 highest ranked national teams using combination of loss ratio and number of games the two teams played as weighting function (function 1 in Table~\ref{tab1}) and 0.05 damping factor. The 4-th column gives their position in the official ranking} 
\begin{tabular}{| c | c | c | c | c |}
    \hline
    \textbf{\#} & \textbf{COUNTRY} & \textbf{PAGERANK} & \textbf{OFFICIAL} \\ \hline 
    1 & Brazil & 0.040375 & \color{green}\textbf{1} \\ \hline 
	2 & Italy & 0.037992 & 3 \\ \hline 
	3 & Germany & 0.033801 & 2 \\ \hline 
	4 & Netherlands & 0.031052 & 8 \\ \hline 
	5 & Argentina & 0.029159 & 4 \\ \hline 
	6 & England & 0.029100 & \color{green}\textbf{6} \\ \hline 
	7 & Spain & 0.027904 & 5 \\ \hline 
	8 & France & 0.025670 & 7 \\ \hline 
	9 & Czechoslovakia & 0.025155 & \color{grey}\textbf{NA} \\ \hline 
	10 & Sweden & 0.022882 & \color{green}\textbf{10} \\ \hline 
	11 & Mexico & 0.022034 & 13 \\ \hline 
	12 & Hungary & 0.022014 & 16 \\ \hline 
	13 & Uruguay & 0.020660 & 9 \\ \hline 
	14 & Belgium & 0.020255 & \color{green}\textbf{14} \\ \hline 
	15 & Portugal & 0.020211 & 17 \\ \hline 
	16 & Poland & 0.019528 & 15 \\ \hline 
	17 & Denmark & 0.019206 & \color{red}\textbf{25} \\ \hline 
	18 & Croatia & 0.018993 & \color{red}\textbf{27}\\ \hline 
	19 & Switzerland & 0.016650 & 21\\ \hline 
	20 & Yugoslavia & 0.016466 & \color{grey}\textbf{NA} \\ \hline 
\end{tabular}
\end{table}

\begin{figure}[h]
\centering
\includegraphics[width=0.45\textwidth]{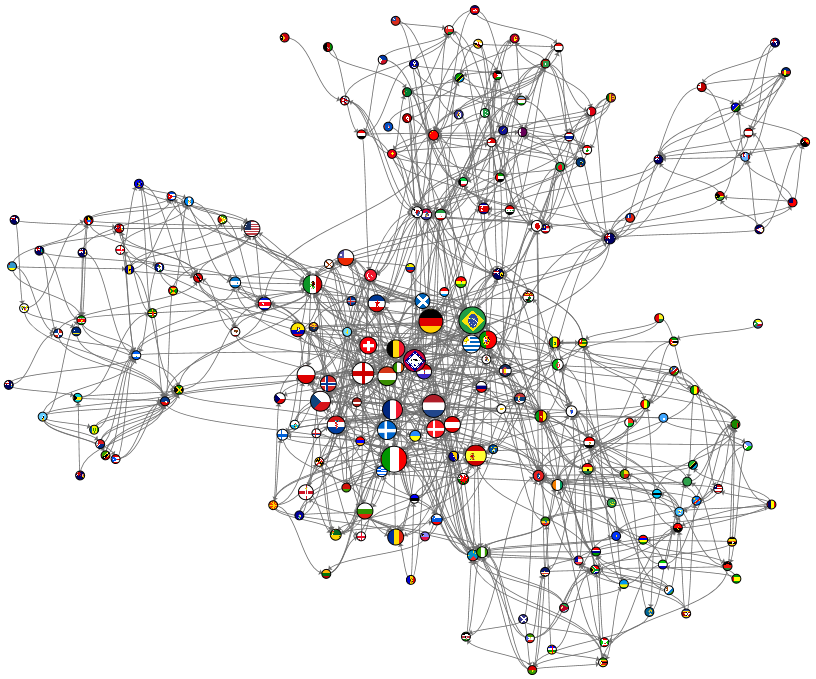}
\caption{\label{fig2}The graph of the match-ups built with the combination of loss ratio and number of games the two teams played as weighting function (function 1 in Table~\ref{tab1}). The size of each node corresponds to their PageRank (damping factor of 0.05 used). For the sake of clarity only the strongest links coming out of each node are shown.}
\end{figure} 

\begin{figure}[h]
\includegraphics[scale=0.43]{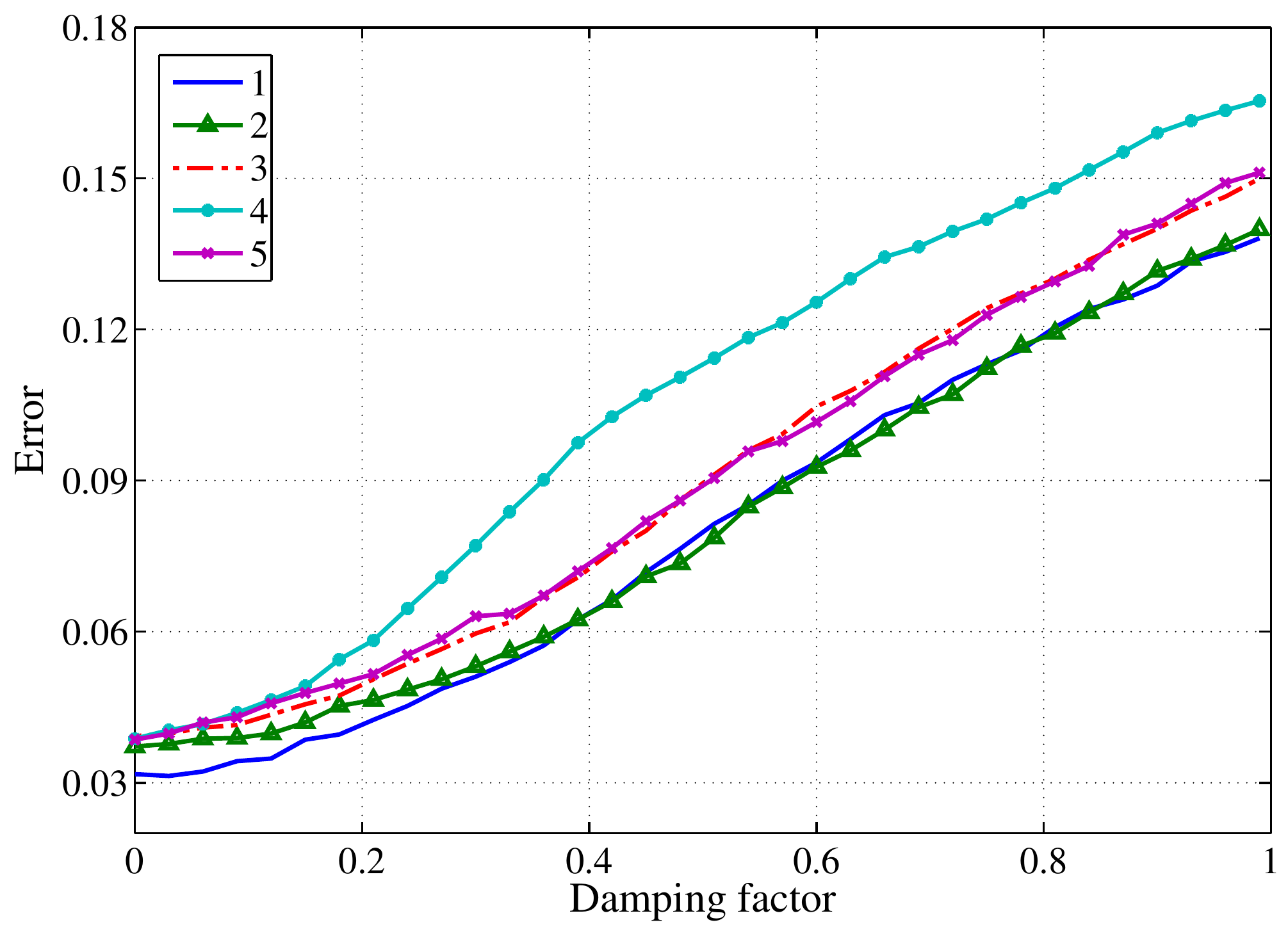}
\caption{\label{fig3} The error in normalized number of inversions of the first 5 weighting functions in Table~\ref{tab1} as function of the damping factor}
\end{figure}

In order to find the most precise ranking several different weighting functions have been tried and almost all of them delivered similar results. The results were evaluated by comparing the PageRank to the official world cup ranking. We have used normalized number of inversions as evaluation metric~\cite{knuth1998art}, taking the official FIFA all-time rankings as referent ordering. The tested weighting functions and their scores are listed at Table~\ref{tab1}. Lower score means the results generated using the corresponding metric are more similar to the official ranking. We only used the top 30 highest ranked teams in the comparison because we wanted to give them higher priority and get their ordering right at the cost of misplacing some of the lower rated teams. The error of the weighting functions also depends on the damping factor. The minimum is achieved when the damping factor value is very small, around 0.05. That is the value we used in the evaluations of the metrics shown in Table~\ref{tab1}. Fig~\ref{fig3} shows errors (in normalized inversions count) for the top 5 metrics as functions of the damping factor. As expected the error increases with the growth of the damping factor. Table~\ref{tab4} shows the top 20 teams (for brevity), according to our best weighting function. The 4-th column contains the positions for each team at the official rankings board. The position is marked green if the team holds the same place in both ours and the official rankings. The position is marked with red if there is a large displacement (Denmark and Croatia). If a team is not found in the official ranking (Czechoslovakia and Yugoslavia in our case) their position is marked with NA. Fig~\ref{fig2} shows the match-ups graph. Each team is a node in the graph represented by their national flag and the size of each node is proportional to it's PageRank. In the figure a portion of the links are omitted for the sake of clarity, thus the real graph is much denser than it appears.  

Possible issue when using PageRank as ranking method might be the following: A node can obtain a high PageRank score if it has a high ranked neighbour from which it can receive significant amount of votes or if it has many low ranked neighbours. In our example, if a national team is high ranked then they must have either defeated many low ranked teams or achieved remarkable results against a highly ranked opponent. This property of the Random Walk affects our results especially since we treat all matches equally, without taking into account whether it is qualification round or final game. As a result there might be teams that have received high ranking only because they have played and won against many low ranked opponents in less significant qualification matches. 

\section{Conclusion}
\label{s4}

Throughout this paper we explored the PageRank method for ranking national football teams. Our results showed that even with simple weighting functions such as ratio of the goals scored or matches won, the PageRank algorithm derives promising results. The rankings this method produced are similar to the official FIFA all-time rankings. However, it is difficult to evaluate whether the PageRank with use of more sophisticated weighting function and more features within the dataset could lead to a better rating scheme than the official. Anyway, under the assumption that the FIFA ranking system is proper and accurate, RandomWalk despite the simple dataset and weighting metrics can replicate it's results in a great deal.


\section*{Acknowledgment}
We would like to thank Andrej Gajduk and Igor Trpevski for fruitful discussions and comments. VL also thanks TAPAN MNG D.O.O.E.L. Negotino for the internship opportunity during which the presented work was completed.


\begin{thebibliography}{10}

\providecommand{\url}[1]{#1}
\csname url@samestyle\endcsname
\providecommand{\newblock}{\relax}
\providecommand{\bibinfo}[2]{#2}
\providecommand{\BIBentrySTDinterwordspacing}{\spaceskip=0pt\relax}
\providecommand{\BIBentryALTinterwordstretchfactor}{4}
\providecommand{\BIBentryALTinterwordspacing}{\spaceskip=\fontdimen2\font plus
\BIBentryALTinterwordstretchfactor\fontdimen3\font minus
  \fontdimen4\font\relax}
\providecommand{\BIBforeignlanguage}[2]{{%
\expandafter\ifx\csname l@#1\endcsname\relax
\typeout{** WARNING: IEEEtran.bst: No hyphenation pattern has been}%
\typeout{** loaded for the language `#1'. Using the pattern for}%
\typeout{** the default language instead.}%
\else
\language=\csname l@#1\endcsname
\fi
#2}}
\providecommand{\BIBdecl}{\relax}
\BIBdecl

\bibitem{fifa-competitions}
\BIBentryALTinterwordspacing
F.~I. de~Football~Association \emph{et~al.}, ``Fifa competitions and olympic
  football tournaments 1908-2017,'' 2014. [Online]. Available:
  \url{http://www.fifa.com/worldcup/organisation/documents/index.html}
\BIBentrySTDinterwordspacing

\bibitem{fifa-comparative}
\BIBentryALTinterwordspacing
------, ``Fifa world cup comparative statistics 1982-2014,'' 2014. [Online].
  Available:
  \url{http://www.fifa.com/worldcup/organisation/documents/index.html}
\BIBentrySTDinterwordspacing

\bibitem{hoffmann2002socio}
R.~Hoffmann, L.~C. Ging, and B.~Ramasamy, ``The socio-economic determinants of
  international soccer performance,'' \emph{Journal of Applied Economics},
  vol.~5, no.~2, pp. 253--272, 2002.

\bibitem{pena2012network}
J.~L. Pe{\~n}a and H.~Touchette, ``A network theory analysis of football
  strategies,'' \emph{arXiv preprint arXiv:1206.6904}, 2012.

\bibitem{duch2010quantifying}
J.~Duch, J.~S. Waitzman, and L.~A.~N. Amaral, ``Quantifying the performance of
  individual players in a team activity,'' \emph{PloS one}, vol.~5, no.~6, p.
  e10937, 2010.

\bibitem{hughes2005analysis}
M.~Hughes and I.~Franks, ``Analysis of passing sequences, shots and goals in
  soccer,'' \emph{Journal of Sports Sciences}, vol.~23, no.~5, pp. 509--514,
  2005.

\bibitem{dixon1998birth}
M.~Dixon and M.~Robinson, ``A birth process model for association football
  matches,'' \emph{Journal of the Royal Statistical Society: Series D (The
  Statistician)}, vol.~47, no.~3, pp. 523--538, 1998.

\bibitem{fifa-coca-cola}
``Fifa/coca-cola world ranking,''
  \url{http://www.fifa.com/fifa-world-ranking/ranking-table/men/}, accessed:
  2015-01-25.

\bibitem{fifa-all-time-rankings}
\BIBentryALTinterwordspacing
F.~I. de~Football~Association \emph{et~al.}, ``Fifa world cup all-time
  ranking,'' 2014. [Online]. Available:
  \url{http://www.fifa.com/worldcup/organisation/documents/index.html}
\BIBentrySTDinterwordspacing

\bibitem{elo}
``World football elo ratings,'' \url{http://www.eloratings.net}, accessed:
  2015-02-11.

\bibitem{page1999pagerank}
L.~Page, S.~Brin, R.~Motwani, and T.~Winograd, ``The pagerank citation ranking:
  Bringing order to the web.'' 1999.

\bibitem{backstrom2011supervised}
L.~Backstrom and J.~Leskovec, ``Supervised random walks: predicting and
  recommending links in social networks,'' in \emph{Proceedings of the fourth
  ACM international conference on Web search and data mining}.\hskip 1em plus
  0.5em minus 0.4em\relax ACM, 2011, pp. 635--644.

\bibitem{kimura2006tractable}
M.~Kimura and K.~Saito, ``Tractable models for information diffusion in social
  networks,'' in \emph{Knowledge Discovery in Databases: PKDD 2006}.\hskip 1em
  plus 0.5em minus 0.4em\relax Springer, 2006, pp. 259--271.

\bibitem{stanoev2011identifying}
A.~Stanoev, D.~Smilkov, and L.~Kocarev, ``Identifying communities by influence
  dynamics in social networks,'' \emph{Physical Review E}, vol.~84, no.~4, p.
  046102, 2011.

\bibitem{erkan2004lexrank}
G.~Erkan and D.~R. Radev, ``Lexrank: Graph-based lexical centrality as salience
  in text summarization,'' \emph{J. Artif. Intell. Res.(JAIR)}, vol.~22, no.~1,
  pp. 457--479, 2004.

\bibitem{agirre2009personalizing}
E.~Agirre and A.~Soroa, ``Personalizing pagerank for word sense
  disambiguation,'' in \emph{Proceedings of the 12th Conference of the European
  Chapter of the Association for Computational Linguistics}.\hskip 1em plus
  0.5em minus 0.4em\relax Association for Computational Linguistics, 2009, pp.
  33--41.

\bibitem{keener1993perron}
J.~P. Keener, ``The perron-frobenius theorem and the ranking of football
  teams,'' \emph{SIAM review}, vol.~35, no.~1, pp. 80--93, 1993.

\bibitem{mukherjee2012identifying}
S.~Mukherjee, ``Identifying the greatest team and captain—a complex network
  approach to cricket matches,'' \emph{Physica A: Statistical Mechanics and its
  Applications}, vol. 391, no.~23, pp. 6066--6076, 2012.

\bibitem{radicchi2011best}
F.~Radicchi, ``Who is the best player ever? a complex network analysis of the
  history of professional tennis,'' \emph{PloS one}, vol.~6, no.~2, p. e17249,
  2011.

\bibitem{11v11}
``11v11 - home of football statistics and history,''
  \url{http://www.11v11.com}, accessed: 2015-01-25.

\bibitem{brin1998anatomy}
S.~Brin and L.~Page, ``The anatomy of a large-scale hypertextual web search
  engine,'' \emph{Computer networks and ISDN systems}, vol.~30, no.~1, pp.
  107--117, 1998.

\bibitem{langville2004deeper}
A.~N. Langville and C.~D. Meyer, ``Deeper inside pagerank,'' \emph{Internet
  Mathematics}, vol.~1, no.~3, pp. 335--380, 2004.

\bibitem{knuth1998art}
D.~E. Knuth, \emph{The art of computer programming: sorting and
  searching}.\hskip 1em plus 0.5em minus 0.4em\relax Pearson Education, 1998,
  vol.~3.

\end{thebibliography}
\end{document}